%% file: main.tex
\documentclass[twocolumn]{aastex63}
\pdfoutput=1 
\usepackage{amsmath,amsfonts,amssymb}
\usepackage{graphicx}
\usepackage{blindtext}
\usepackage[T1]{fontenc}
\usepackage{apjfonts} 
\usepackage[figure,figure*]{hypcap}

\input{mycommands}

\shorttitle{The Discovery of A Disk of Dark Molecular Gas}
\shortauthors{Busch et al.}

\draft

\begin{document}

\title{Observational Evidence for a Thick Disk of Dark Molecular Gas in the Outer Galaxy}

\author[0000-0003-4961-6511]{Michael P. Busch}
\thanks{To whom correspondence should be addressed: mpbusch@jhu.edu}
\affiliation{Department of Physics and Astronomy, Johns Hopkins University, 3400 North Charles Street, Baltimore, MD 21218, USA}
\affiliation{NSF GRFP Fellow}
\affiliation{LSSTc DSFP Fellow}
\author[0000-0003-3821-5069]{Philip D. Engelke}
\affiliation{Department of Physics and Astronomy, Johns Hopkins University, 3400 North Charles Street, Baltimore, MD 21218, USA}
\affiliation{Frontier Technology Inc., 100 Cummings Center Suite 450G, Beverly, MA, 01915, USA}
\author{Ronald J. Allen}
\thanks{Deceased as of the date of publication.}
\affiliation{Department of Physics and Astronomy, Johns Hopkins University, 3400 North Charles Street, Baltimore, MD 21218, USA}
\affiliation{Space Telescope Science Institute, 3700 San Martin Drive, Baltimore, MD 21218, USA}
\author{David E. Hogg}
\affiliation{National Radio Astronomy Observatory, 520 Edgemont Road, Charlottesville, VA 22903, USA}

\begin{abstract}

We present the serendipitous discovery of an extremely broad ($\Delta V_{LSR} \sim 150$ \kmps), faint ($T_{mb} < 10 \textrm{mK}$), and ubiquitous 1667 and 1665 MHz ground-state thermal OH emission towards the 2nd quadrant of the outer Galaxy ($R_{gal}$ > 8 kpc) with the Green Bank Telescope. Originally discovered in 2015, we describe the redundant experimental, observational, and data quality tests of this result over the last five years. The longitude-velocity distribution of the emission unambiguously suggests large-scale Galactic structure. We observe a smooth distribution of OH in radial velocity that is morphologically similar to the HI radial velocity distribution in the outer Galaxy, showing that molecular gas is significantly more extended in the outer Galaxy than previously expected. Our results imply the existence of a thick ($-200< z < 200$ pc) disk of diffuse ($n_{H_{2}}$ $\sim$ 5 $\times$ 10$^{-3}$ cm$^{-3}$) molecular gas in the Outer Galaxy previously undetected in all-sky \twCO\ surveys.

\end{abstract}


\section{Introduction}

Molecular Hydrogen (\Htwo) is the most abundant molecule in the Universe \citep{Draine2011PhysicsMedium}. It is important for cooling of the interstellar medium and thus regulates star-formation in galaxies \citep{Glover2008UncertaintiesFormation}. It is a catalyst for interstellar chemistry. \Htwo\ can be observed using the higher J rotational states in warm molecular gas \citep{Goldsmith2010MOLECULARCLOUD}, but these require energies usually not available in cold molecular gas, where the bulk of the \Htwo\ lies. This means that despite the importance of \Htwo\ to most areas of astrophysics, it remains difficult to study.

The carbon monoxide (CO) molecule has been used historically to trace the overall distribution of molecular gas in our Galaxy and other galaxies \citep{Bolatto2013TheFactor}, due to its substantial relative brightness and abundance. Surveys for the lowest rotational transition of \twCO\ (J=1-0) at 2.6 mm have made invaluable contributions to our knowledge of the overall distribution, morphology and mass content of molecular gas \citep{Heyer2015MolecularWay}. 

There remain difficulties in using CO as a tracer for the total \Htwo\ content. The CO line is commonly optically thick, and thus an uncertain and indirect conversion factor, X(\textrm{CO}), is used to convert between a CO line intensity to a \Htwo\ column density. Secondly, a growing body of observational research using indirect total gas tracers (Dust, $\gamma$-ray) have shown that there exists a significant portion of \Htwo\ in diffuse regions that are not observed in CO surveys \citep{Blitz1990MolecularCO,Grenier2005UnveilingNeighborhood,Ackermann2010ConstraintsQuadrant, PlanckCollaboration2011PlanckGalaxy}. \cite{Wolfire2010TheGas} coined the term `dark molecular gas' to describe the gas not observed by CO, and theorized that up to 30\% of \Htwo\ can be missed by CO in the surface layers of molecular clouds. 

The physical conditions of the dark gas have been mysterious for much of the past decade, but recent research has shown that the missing portion of gas is not predominately explained by optically thick HI \citep{Murray2018OpticallyISM}, which was a promising possible avenue for explaining the dark gas \citep{Fukui2015OpticallyGas}. The indirect tracers of gas content such as dust emission and $\gamma$-rays are suggestive that the content is in fact molecular gas not traced by CO, but only a molecular tracer could provide such confirmation.

Additionally, these tracers cannot provide radial velocity measurements and hence we lack structure information about the distribution of the dark gas in the plane of the Galaxy. Using a molecule, and particularly for chemical reasons, an abundant hydride \citep{Neufeld2016TheClouds} such as CH \citep{Weselak2010TheMolecules, Xu2016CHTAURUS, Jacob2019FingerprintingDeconvolution, Jacob2020FirstMedium} or OH \citep{Weinreb1963RadioMedium,Weinreb1965ObservationsEmission, Heiles1968NormalClouds,Dickey1981Emission-absorptionobservationsClouds, Wannier1993WarmObservations, Allen2012Faint5circ, Dawson2014SPLASH:Region, Allen2015The+1deg,Xu2016EVOLUTIONTAURUS,Engelke2018OHW5, Nguyen2018Dust-GasISM, Rugel2018OHTHOR, Engelke2019OHW5, Busch2019TheArm} can provide both the unambiguous conclusion that the dark gas is indeed diffuse molecular gas, and the radial velocity information from a spectral line to describe the Galactic distribution of the gas by utilizing the velocity field of the Galaxy \citep{Brand1993TheGalaxy.} Studies of the large-scale distribution of gas have been completed mainly using HI \citep{Kalberla2005TheHI, Peek2011The1, Winkel2015TheRelease} or CO \citep{Heithausen1993AQuadrant., Heyer1998TheGalaxy,Dame2001TheSurvey, Heyer2015MolecularWay}, and thus the diffuse (or ``dark``) molecular gas, and its mass content and Galactic structure, are inherently excluded by these studies \citep{Blitz1990MolecularCO, Goldsmith2010MOLECULARCLOUD, Li2015QuantifyingGas, Xu2016EVOLUTIONTAURUS, Li2018WhereDMG}.

The situation also has been uncertain in the outer Galaxy, where an analysis of the diffuse Galactic $\gamma$-ray distribution from the \textit{Fermi} LAT satellite has suggested that ``vast amounts of missing gas in the outer Galaxy are also possible'' \citep{Ackermann2010ConstraintsQuadrant}. This is called the `CR (cosmic ray) gradient problem`, as the observed decline of CR emissivity is inconsistent with CR propagation models. The presence of a warm ($T \sim$ 100-200 K), extremely diffuse ($n \sim$ 5-20 cm$^{-3}$) phase of molecular gas at large galactic radius was suggested by \citet{Papadopoulos2002MolecularDistances} as an explanation for several observational facts.

These findings led us to the conclusion that an additional large-scale molecular gas tracer is needed besides CO, one that could provide an independent estimate of molecular gas content without being subject to the same set of properties and deficiencies as signals from the CO molecule. Desirable features of such a reliable, alternative molecular tracer would include being optically thin, and containing low critical densities for excitation, so as to better detect lower density, more diffuse molecular gas not traced by CO. Such a tracer would provide information on column density, structure, and kinematics, as well as possibly other physical properties of the gas \citep{Busch2019TheArm, Engelke2020Star-formingObservations, Petzler2020TheRegions}.

The OH molecule, first observed in radio by \cite{Weinreb1963RadioMedium}, is one promising alternative molecular tracer. It produces four optically thin spectral lines centered near 18 cm, with critical densities three orders of magnitude below those of the CO(1-0) transition. As such, OH 18 cm emission lines provide a tracer for molecular gas including diffuse, low density regions, and can yield direct calculation of column densities using equations of radiative transfer. 

Since 2012, we have demonstrated in a series of papers \cite[][]{Allen2012Faint5circ,Allen2013ERRATUM:97,Allen2015The+1deg,Busch2019TheArm} that OH successfully functions as an alternate tracer for molecular gas in observations towards a quiescent region of the Outer Galaxy; further observations in the vicinity of the W5 star-forming region extend the use of OH as an alternate tracer to star-forming regions and further develop the techniques of the field \cite[][]{Engelke2018OHW5,Engelke2019OHW5,Engelke2020Star-formingObservations}. These projects confirmed the viability of OH as a tracer for molecular gas and studied the structure, column density, volume density, and kinematics of molecular gas in these regions.

We proposed another project in 2014 for the Robert C. Byrd Green Bank Telescope (GBT), aimed specifically at the Perseus Arm towards the Outer Galaxy, with the intention of analyzing several outstanding questions in the field of Galactic structure from the standpoint of molecular gas as traced by OH. During the course of this observing run, an observation at $l = 108.0^{\circ}$, $b = 3.0^{\circ}$ performed by RJA in September 2015 and reduced and analyzed by PDE, showed evidence of a broad, low bump-like feature between the narrower, taller features associated with the Perseus Arm and other clumps of molecular gas, and most notably visible near $\sim -95$ km/s. The feature did not go away when baseline fitting procedures were performed, and at first we contemplated the possibility that the feature was an observation of the Outer Arm. However, the feature was not amenable to Gaussian fits, and comparisons to individual peaks in the HI spectrum at that coordinate, which typically contain corresponding features, did not match. The initial conclusions were that the spectral feature was not Gaussian in shape, and did not correspond to any specific HI peaks, or that there was a source of error in the analysis. More observations were planned and carried out on the coordinate to improve the signal-to-noise on the detection during the fall of 2015 as well as observations at adjacent coordinates to see if the feature appeared in them or not. The result was an unambiguous broad feature between the spectral peaks, which continued to evade subtraction by baseline fitting procedures. At this point we hypothesized that the feature could be a baseline artifact resulting from reflections in the structure of the GBT.

Further work was carried out beginning in 2018, when MPB and RJA performed observations with the 20m telescope at the Green Bank Observatory (GBO) \footnote{The Green Bank Observatory is a facility of the National Science Foundation operated under cooperative agreement by Associated Universities, Inc.} at the same coordinates to determine whether the broad spectral feature was an artifact of the GBT or a real astronomical detection, and we developed several redundant methods to analyze the baselines attempting to demonstrate that the feature could be an artifact. Nevertheless, these observations and analyses have bolstered the evidence for a detection of a broad, diffuse OH signal in between the arms of the Galaxy. From here, we performed new observations at several other pointings in the vicinity to map out the extent of this broad Galactic feature in Galactic longitude; the full results of that survey will appear in a follow-up paper.


In this paper, we present the serendipitous discovery of extremely broad, faint OH emission towards the Outer Galaxy, seemingly underlying the narrower ($\Delta V_{LSR} =$ 10 \kmps), brighter ($T_{mb} >$ 5mK) line emission from discrete molecular clouds in spiral arms. The velocity extent and morphology suggest that the \Htwo\ content traced by the OH is in the form of extremely diffuse molecular gas in the geometry of a thick disk. We compute the profile integrals in HI and OH and transform those to column densities under a number of assumptions. We then compute the abundance ratio OH/HI, and, using dust reddening as a calibration for the total column density of gas in this direction, we calculate the abundance ratio of OH/\Htwo. Under several assumptions, we also calculate preliminary mass estimates.

\section{Observations and Data}\label{Observations}

In this study, we present observations of the OH molecule at the main line transitions of 1667 and 1665 MHz using two telescopes, the 100m Green Bank Telescope and the 20m Telescope, at the GBO. We utilize the ``SFD`` (Schlegel, Finkbeiner, Davis) dust map \citep{Schlegel1997MapsForegrounds,Schlafly2010MeasuringSFD} through a query from the Python package \textit{dustmaps} \citep{M.Green2018Dustmaps:Dust}.

\subsection{100m Green Bank Telescope}\label{gbt}

We have observed the ground state, $\Lambda$-doubling OH emission lines at 18 cm with the 100m Robert C. Byrd Green Bank Telescope, located in Green Bank, West Virginia, with multiple observational projects over the past five years in order to probe the characteristics of the OH emission from diffuse molecular gas in the outer Galaxy. The project IDs: AGBT14B\_031A, AGBT14B\_031B, AGBT15B\_004, and AGBT19A\_453 were all observed for various OH emission projects in the outer Galaxy, but later were reanalyzed to probe the robustness of the result presented. The first profile showing the broad emission feature was under Project ID AGBT15B\_004 at $l=108^{\circ}$, $b=3^{\circ}$, and was thought to be a `baseline feature` that was difficult to subtract. 

A follow-up Project, AGBT19A\_484, was undertaken during the summer of 2019 with the GBT in order to directly observe the source geometry of the extremely broad OH emission as a function of Galactic longitude and latitude for the first time. These observations were completed by using frequency-switching by 2 MHz centered on 1617, 1665, and 1720 MHz \citep{ONeil2002SingleWavelengths} using the VEGAS spectrometer in the L-band range \citep{Prestage2015ThePlans}. The observations cover all four ground-state OH emission lines, although the 1612 band is commonly unusable due to radio frequency interference (RFI). The beam size of the GBT at 18 cm is 7.6'. 

\subsection{20m GBO Telescope}\label{20m}

After an initial discovery in 2015 in a GBT spectrum towards $l=108.000^{\circ},  b=3.000^{\circ}$, we hypothesized that the broad signal was a faint baseline artifact caused by multipath reflections in the structure of the GBT itself and endeavored to prove that the OH signal was erroneous. This type of instrumental artifact was discovered during the commissioning of the GBT \citep{Fisher2003InvestigationTelescope}, where ripples affect the Y linear polarization more than the X linear polarization due to the geometry of the telescope structure.

To test this hypothesis we observed the same position with the 20m telescope at the GBO, and an additional position separated by a degree on the sky. The 20m telescope has completely different structural reflections from the GBT due to its geometry. The detection of the broad, faint OH emission with two different telescopes eliminated the possibility that the broad OH signal was a systematic inherent in the 100m dish of the GBT.

We observed the extended OH emission with the 20m telescope at the Green Bank Observatory during the summer of 2018. Observing time on the 20m telescope for this project was purchased using funds from the Director's Discretionary funds at Space Telescope Science Institute (STScI). The telescope operated in the L-band (1.3-1.8 GHz) range and provided two spectral windows with one centered on the 1667 and 1665 MHz lines, and the other on the 1720 MHz satellite line. The beam size of the 20m at 18 cm was 45'. While the observations with the GBT used frequency switching On/Off observations, the stability of the 20m receiver allowed for total power observations. 

\subsection{Dust and HI}\label{dustData}

To trace the total gas column density $N_{H}$, we use the all-sky dust reddening map from \citet{Schlegel1997MapsForegrounds, Schlafly2010MeasuringSFD}; the well-known ``SFD`` dust map. We integrate the entire sightline towards the outer Galaxy, hence the more updated 3D dust maps from \citet{Green2017MeasuringApproach} are not an appropriate estimation of the total gas column due to the lack of stellar measurements a few kpc outside the solar circle. Therefore, we opt to use the 2D SFD dust map. This map has an angular resolution of 6'.1, which is comparable to the GBT beam ($\sim$ 7'.8). We query the 2D SFD dustmap for the coordinates $l=108.000^{\circ},  b=3.000^{\circ}$, which returns a dust reddening value, E(B-V) of 1.86 mag.

We also use HI data to estimate the atomic hydrogen column density N(H). We use the HI profile towards $l=108.000^{\circ},  b=3.000^{\circ}$ from the Leiden-Argentine-Bonn (LAB) survey \citet{Kalberla2005TheHI} and HI4PI database \citep{BenBekhti2016HI4PI:GASS}. Additionally, for an estimate of the optical depth of HI towards this sightline, needed for an accurate estimate of N(H), we use two HI absorption-emission pairs from \citet{Strasser2007Absorption}, to calculate the median HI optical depth along the entire line of sight.

\section{Data Reduction}\label{data}

The GBT and 20m OH data were reduced using the \textit{GBTIDL} software \citep{Garwood2006GBTIDL:Data}. The quality of this data is generally very high, and each polarization of each 10 minute scan was reviewed for the presence of radio frequency interference and problems from instrumental effects. In this section, we review the experimental and observational tests of the broad OH emission that heightened our confidence in the result.

\subsection{Separating the ``ON'' and ``OFF'' Spectrum}\label{frequencyswitching}

\begin{figure}[ht!]
        \centering
        \includegraphics[width=0.5\textwidth]{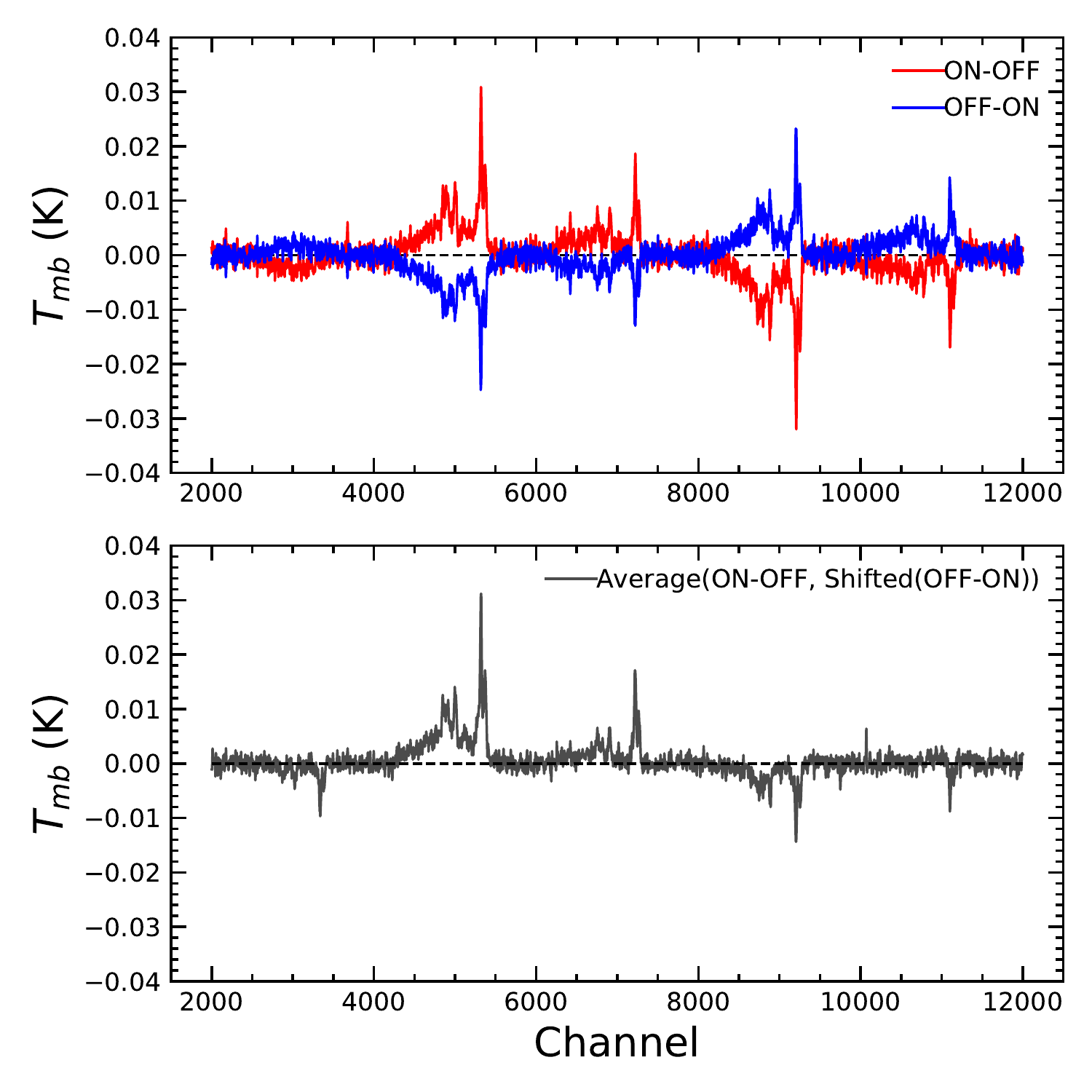}
    \caption{The top panel is the ``ON-OFF'' phase and ``OFF-ON'' phase plotted separately. The second spectrum is a shift and average of the top two spectra, typical of frequency-switched measurements \citep{ONeil2002SingleWavelengths}.}
    \label{fig:frequencyswitching}
\end{figure}

The observations in \textit{AGBT15B\_004} were experimented on in several stages of the frequency switching process in order to test different parts of the intermediate frequency (IF) system to make sure the signal was not introduced by some previously unknown systematic that was unaccounted for. Our observations are completed with in-band frequency switching \citep[see e.g.][]{ONeil2002SingleWavelengths}; the reference measurement of ON-OFF is completed by shifting the central frequency 2 MHz and differencing the spectrum. When we perform in-band frequency switching, the science signal is still present in the ``OFF'' measurement, and the final reduced signal is compiled by shifting and averaging the ``OFF-ON'' and ``ON-OFF'' measurements to decrease the noise in the final RMS by $\sqrt{2}$. This process is demonstrated in Fig. \ref{fig:frequencyswitching}. 

In each separate ON-OFF and OFF-ON signal, we should expect two signals from the two OH lines, and the so-called ``ghosts'' of the subtracted signals from the shifted frequency. Whichever phase is ``ON'' or ``OFF'' is arbitrary. However, it is important that we check if the broad feature is introduced in one part of the IF by appearing in one of the ``phases'' and not the other. If it were to appear in ``ON'' but not ``OFF'', this would raise the alarm of an unknown systematic. We show in Fig. \ref{fig:frequencyswitching} that this is not the case, and the positive and negative signals of the OH emission are clearly visible in both reference spectra. The final product of the frequency switched measurement is shown in the bottom panel of Fig. \ref{fig:frequencyswitching}, where the ``OFF-ON'' measurement is shifted and averaged together with the ``ON-OFF'' measurement. 

In this final spectrum, we would expect the two signals as well as four `ghosts', or negative, subtracted signals. Higher order polynomial fits are required for such wide, frequency-switched measurements, as opposed to the linear baseline fits typically used in position-switched measurements.

\subsection{Baseline Fitting}\label{baseline}

We were immediately suspicious that the broad OH signal was an erroneous signal introduced by high-order polynomial baseline fitting. We performed baseline fits for a wide range of polynomial orders to test the robustness of this method. The broad emission feature is detected in spectra with polynomial baseline fits from order 8 - 23. The precision of the measurement of the profile integral from the GBT spectra is then constrained by the fitting procedure, which introduces $\sim$ a few percent \textit{systematic} error. To test the robustness of the data pipelines; we created three different data pipelines utilizing different common signal reduction techniques including: Fourier filtering (or bandstop filtering), ridge regression (or L2 Regularization), and the standard general orthogonal polynomial regression used in \textit{GBTIDL}. None of the data pipelines generated different results.


In the case of our GBT observations, since the emission feature is very wide in velocity, and the data are frequency-switched, \textit{linear baselines cannot be removed}. A new data fitting procedure in \textit{GBTIDL} had to be created to deal with such wide profiles. Discontinuous regions of the bandpass that do not have `ghosts' (negative signal) introduced by the frequency-switching, or signal from the 1667 and 1665 MHz feature, were used ($N_{chans} \sim 6000$) to fit a single high-order polynomial ranging from order 17-23 based on a few different criteria: a) there should be no positive velocity emission due to galactic structure, b) the chosen polynomial fit does not preferentially subtract or under-subtract signal from the 1667 or 1665 emission near local-gas velocities, resulting in a flat baseline, and c) the rms in the highlighted baseline regions is minimized until no further improvement occurs. This procedure produced the most consistent results for the GBT spectrum and flat baselines in all regions where no signal or `ghost' were expected.

The pipelines were also used on the 20m spectra to produce the broad OH signal with identical results to the GBT spectra. The only noticeable difference in the spectra was the `flattening' of `spiral arm' emission from discrete clouds due to beam dilution. The size of the beam at 18 cm for the GBT is 7.6', whereas the size of the beam for the 20m is approximately 45'. While the surface brightness is larger in the larger beam, the similar surface brightness of the broad OH emission feature in both spectra indicate that the filling factor of the emitting gas is high, suggesting that the source geometry of the OH emission is extremely diffuse molecular gas that is volume filling.

In addition to replicating the GBT result with the 20m telescope, we also opted to forgo frequency-switching so that the baseline fitting of the 20m spectra would be less complex. Baseline fits of order 2-3 were fit to the total power spectrum and clearly showed the broad OH feature. This result demonstrated that detection of the broad feature does not rely on the higher order (i.e. 17-23) polynomial fits, which conceivably could have introduced spurious artificial signals into the GBT spectrum. In the 100 hour 20m OH spectrum, the RMS noise is $\sim$ 0.3 mK at a velocity resolution of 1 \kmps.

\subsection{Separating the Linear Polarization}\label{polarization}

\begin{figure}[ht]
    \centering
    \includegraphics[width=0.5\textwidth]{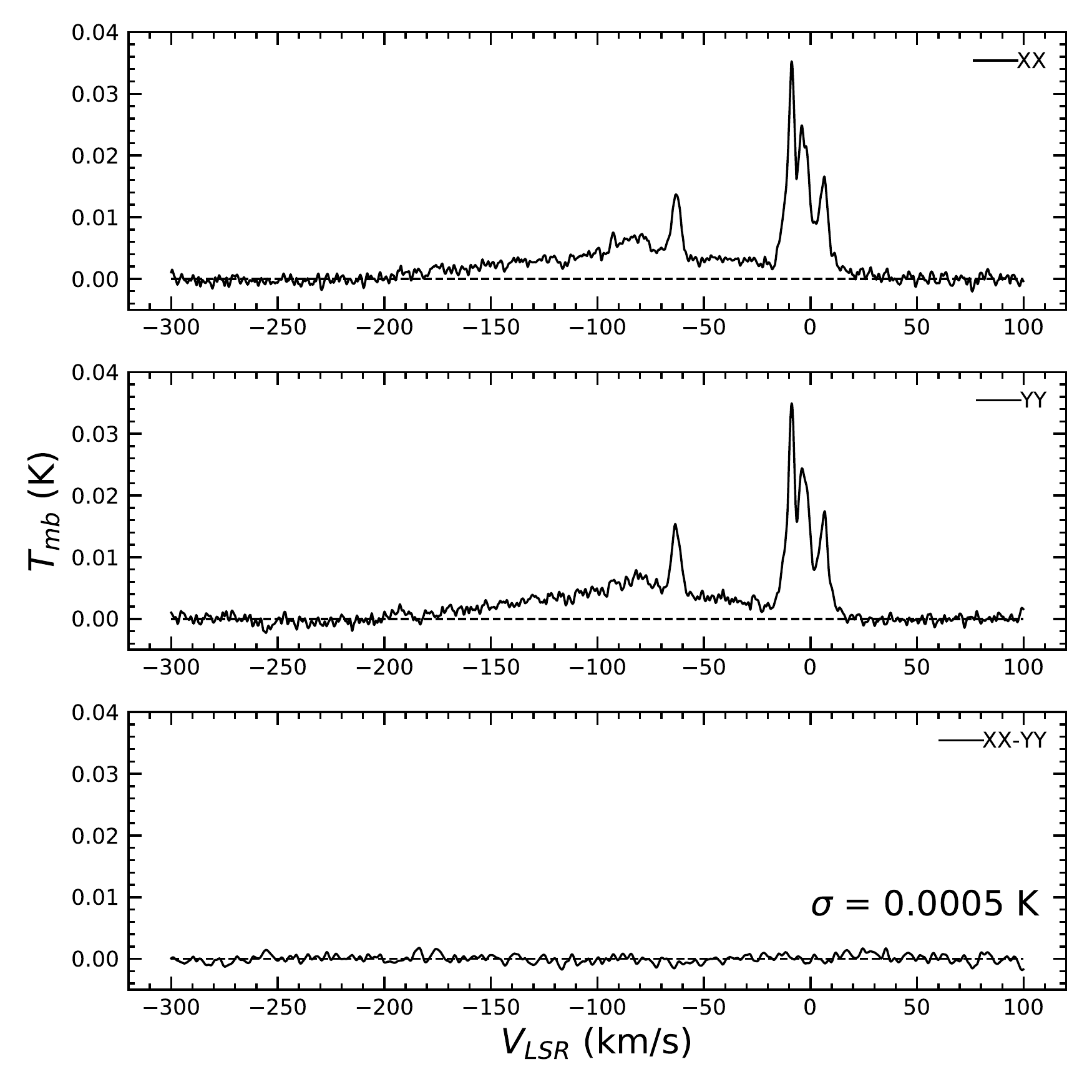}
    \caption{The broad OH emission feature as seen in the GBT spectra. Top Panel: A stacked 40 hr spectrum of XX linear polarization. Middle Panel: A stacked 40 hr spectrum of YY linear polarization. Bottom Panel: The resulting residual spectrum from XX-YY. The residual noise is possibly from an additional known spectral ripple in the YY polarization due to the geometry of the GBT itself \citep{Fisher2003InvestigationTelescope}.}
    \label{fig:residual}
\end{figure}

The emission appears to be unpolarized. This is an important check because at cm wavelengths at the GBO observers will sometimes experience weak, transient and usually polarized radio frequency interference (RFI) that is time-dependent. We expect any real thermal OH emission to be non-polarized because it is due to particle motions and collisions in gas that result in dipole oscillations among the four ground-state lines. We checked that we could reproduce the characteristics of the broad and faint OH emission features in 1667 and 1665 MHz in the independent orthogonal linear polarizations (XX and YY). While the YY polarization experiences more baseline noise due to the aforementioned multi-path scattering, we were able to replicate the spectrum and calculate a residual spectrum between the polarization which shows a spectrum with zero emission from the velocity range of the broad OH feature. If the emission were only detected in one polarization, that would indicate that it would be some type of systematic error that we were previously not accounting for. 

In this experiment, we stacked five GBT OH spectra towards $l = 108.000^{\circ}$, $b = 3.00^{\circ}$, keeping the linear polarizations separate throughout the reduction. Then, the spectra are smoothed to a velocity resolution of 1 \kmps. The XX and YY spectra are displayed in the top two panels of Fig. \ref{fig:residual}. The residual spectrum of XX-YY is also shown to demonstrate that the two polarizations behave differently at a very small level. The resulting integration time on the spectra in Fig. \ref{fig:residual} is approximately 40 hours with a RMS noise of $T_{RMS}$ = 300 $\mu$K. The systematic from the YY spectrum is at the level of 0.5-0.6 mK, the brightness temperature of the 1667 MHz emission is higher at $T_{mb} = 1-3 mK$, and the 1665 MHz emission is about half of that, allowing a detection of this feature at high S/N, especially if only using XX data. We rule out the possibility that the emission from the this feature is a polarized systematic baseline ripple in the GBT.

\section{Results}\label{results}

In this section we will describe the two main spectra presented in this paper, the nature of the OH main lines, and how column densities of HI, OH and \Htwo\ are computed.

\begin{figure*}[ht!]
        \centering
        \includegraphics[width=0.85\textwidth]{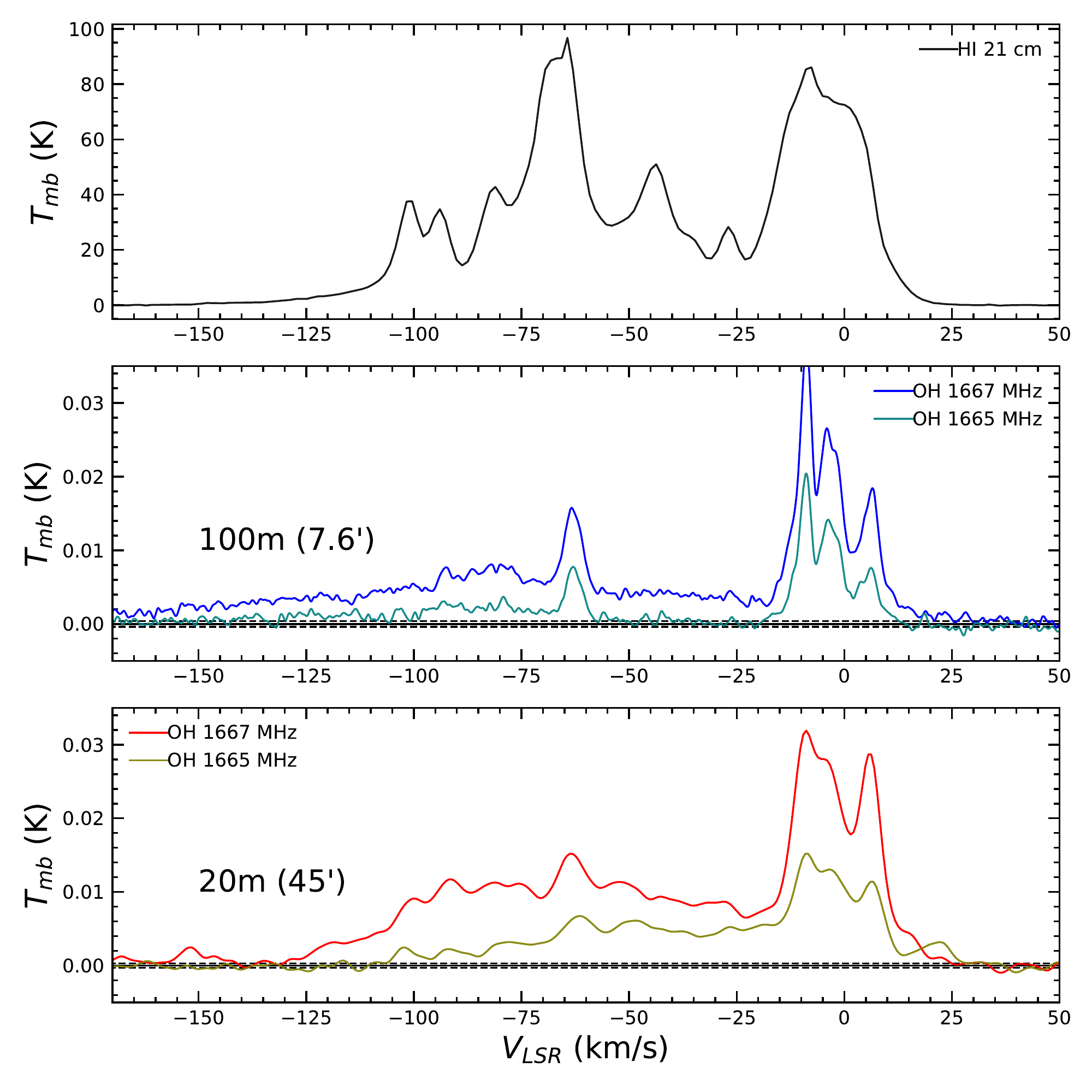}
    \caption{The observed broad emission in spectra towards the Outer Galaxy at $l = 108^{\circ}$, $b = 3^{\circ}$ comparing the results from the 100m GBT and the 20m telescope at the GBO. The top panel shows the HI emission at 1420 MHz (21 cm) observed with the GBT. The middle panel shows the OH spectra observed at 1667 MHz (blue) and at 1665 MHz (green) using the 100m GBT, while the bottom panel shows the OH spectra observed at 1667 MHz (red) and 1665 MHz (yellow-brown) using the 20m telescope at the GBO. Note that while the telescope resolution varies from 7.6' at the GBT to 45' at the 20m telescope, the broad emission is detected with both of these telescopes. The dashed lines are the $3\sigma$ statistical detection limit, corresponding to 3 mK for the 100m telescope and 2 mK for the 20m telescope.}
    \label{fig:allcombined}
\end{figure*}

\subsection{GBT Spectra}\label{gbtspectra}

It became apparent that many sightlines could be stacked from multiple OH projects because of the extended nature of the broad OH emission, which appeared to be beam filling. This allowed for artificially extremely long integration times (up to 300 hours). Depending on how many positions are averaged together, the rms noise in the spectra was between $\sim$ 300-800 $\mu$K. The GBT spectra shown in the middle panel of Fig. \ref{fig:allcombined} come from an 80 hour, stacked OH spectrum centered at 1666 MHz, which was then re-centered for the 1667 and 1665 MHz lines respectively. The positions used in this spectra are seven ten-hour exposures around the coordinates ($l=108\degr$, $b=3\degr$). The main beam efficiency of the 100m at L-band is estimated to be $\eta_{mb}$ = 0.95, which we used to convert antenna temperature to main beam temperature.

\subsection{20m Spectra}\label{20mspectra}

After averaging 100 hrs of observations at two separate positions ($l=108\degr$, $b=3\degr$ and $l=109\degr$, $b=2.5\degr$), separated by a degree on the sky, and smoothing to 1 \kmps, the rms noise in the spectrum was $\sim$ 300 $\mu$K. We have aligned the local gas emission in the two spectra using a LSR (local standard of rest) velocity correction calculator kindly provided by F. Ghigo\footnote{https://www.gb.nrao.edu/~fghigo/gbt/setups/radvelcalc.html}. The resulting OH spectrum at 1667 and 1665 MHz are shown in the bottom panel of Fig. \ref{fig:allcombined}. The main beam efficiency of the 20m is estimated to be $\eta_{mb}$ = 0.77 (F. Ghigo, private communication), which we used to convert antenna temperature to main beam temperature.

\subsection{1667 and 1665 MHz OH Lines}\label{OHlines}

On a GBT spectrum with 10 hours of integration time, the resulting rms at a velocity resolution of 1 \kmps\ is close to 1 mK. Typical brightness temperatures per channel are $T_{mb} \sim 1-3$mK, and thus the profile integral over the large velocity width ($\Delta v_{LSR} \sim 150 \kmps$) of the emission results in a signal-to-noise (S/N) of approximately 25. The emission also appears clearly in the two main lines of the OH ground state ($\Delta F = 0$) at 1667 and 1665 MHz. The profile integrals of the observed emission regularly appear to be in the 9:5 LTE ratio set by the quantum mechanical transition strengths \citep{Townes1955MicrowaveSpectroscopy, Turner1971NonthermalClouds}. This ratio indicates that collisional excitation of OH is the dominant radiation mechanism in the gas, as the energy levels approach the typical Boltzmann distribution as described by an excitation temperature $T_{ex}$, yet the brightness temperature is not near the actual gas temperature.

\subsection{Column Density of OH} \label{OH}

The expression for the total molecular gas column density is \citep{Mangum2015HowDensity}:

\begin{equation}
    N_{tot} = \frac{3h}{8 \pi^{3} |\mu_{lu}|^{2}} \frac{Q_{rot}}{g_{u}} \exp\left(\frac{E_{u}}{kT_{ex}}\right) [\exp\left(\frac{h\nu}{kT_{ex}}\right)-1]^{-1} \int \tau_{\nu}d\nu
\end{equation}

where $Q_{rot}$ is the rotational partition function, the statistical sum over all rotational energy levels in the molecule (for the OH hyperfine, $\Lambda$-doubling ground states, this sum is 16), and $g_{u}$ is the degeneracy of the energy level $u$.

In the analysis of OH emission line profiles, this equation can be converted to mostly observables. Consider a molecular cloud located in the ISM at a distance S along a line of sight from the Sun towards the outer Galaxy. The beam-averaged column density of OH molecules $\left<N(OH)\right>$ along the line of sight through the cloud is \citep{Liszt2010TheGas}:

\begin{equation}
    \left<N(\mathrm{OH})\right> = C_{67}\left[\frac{T_{ex}^{67}}{T_{ex}^{67} - T_c}\right]
    \int \Delta T_b^{67}(v) dv ;
    \label{eqn:coldens}
\end{equation}

where $\Delta T_b^{67}(v)$ is the (main beam) brightness temperature of the OH emission profile from the cloud as observed in one of the 18-cm OH transitions (in this example the transition that gives rise to the 1667 MHz line), minus an estimate of the underlying radio continuum brightness at the same radial velocity (the `baseline'); $T_{ex}^{67}$ is the excitation temperature for the 1667 line; $T_c$ is the brightness of the Galactic continuum emission at 1667 MHz incident on the back surface of the cloud, and the integration over velocity includes all the molecular emission thought to arise in that particular cloud. This equation assumes that the Rayleigh-Jeans approximation applies, that the optical depth of the cloud in the OH line is small, that all the OH molecules are in the ground rotational-vibrational state, and that the 4 ground-state levels are populated according to LTE. Under these conditions, the constant $C_{\nu}$ consists of:
\begin{equation}
    C_{\nu} = \frac{8 \pi k \nu_{ul} \Sigma g_i}{hc^2 A_{ul}g_u} \times (\nu_{ul}/c) 
\end{equation}

where $\nu_{ul}$ is the transition line frequency, $A_{ul}$ is the Einstein A coefficient for that transition, $\Sigma g_i$ is the sum of the statistical weights of the four levels giving rise to the 18cm lines (=16), $g_u$ is the statistical weight of the upper energy level that gives rise to the specific line (e.g.\ = 5 for the 1667 line), and k, h, and c designate the usual physical constants. This equation has been grouped into two terms; the first is from  \cite{Goss1968OHGalaxy}, and the second provides the conversion from line widths in Hertz (used by Goss) to the currently more conventional units of doppler velocity. Using the entries for the line frequencies and Einstein A coefficients found in the Splatalogue data base\footnote{http://www.cv.nrao.edu/php/splat/} appropriate for the 1667 MHz line, the value of $C_{67}$ is $2.257 \times 10^{14}$ \pcmsq\ for $\Delta T_b^{67}$ in Kelvins and velocity in \kmps.

In this analysis, we assume an excitation temperature of 5.1K $\pm$ 1K, as the excitation temperature is usually found to be within 1-2K above the continuum temperature \citep{Li2018WhereDMG,Engelke2020Star-formingObservations}, although we stress caution because the real excitation temperature of the gas may be lower, closer to the Galactic background as found in \citet{Li2018WhereDMG}. The possible range of excitation temperatures and continuum temperatures alone introduces systematic uncertainty in the resulting N(OH) values, which we explore below. To obtain an estimate of the continuum background brightness temperature, $T_{c}$, we estimate the synchrotron contribution from the 408 MHz continuum map of the Canadian Galactic Plane Survey (CGPS, \citet{Taylor2003TheSurvey}) by adopting a temperature spectral index of 2.8 \citep{Reich1988ASky.}:

\begin{equation}
    T_{c} = 2.7 + T_{c, 408}(1667/408)^{-2.8},
\end{equation} 

which results in a T$_{c}$ of approximately 4K. An approved GBT program for this coming year will attempt to observe this dark disk in absorption and emission pairs, which should unambiguously provide an accurate measurement of $T_{ex}$ in the future. The resulting column density is: $N($OH$)$ = (7.41 $\pm$ 0.17) $\times$ 10$^{14}$ cm$^{-2}$, which includes statistical error. For a sensitivity calculation on systematic errors, we calculate the possible ranges of OH column densities using 4.1K < $T_{c}$ < 4.5 K, and 4.6K < $T_{ex}$ < 6.1 K. This range results in: $N($OH$)$ = (1.0 $\pm$ 0.017 statistical $\pm$ 0.4 systematic) $\times$ 10$^{15}$ cm$^{-2}$. Note that the systematic uncertainty due to the unknown values of $T_{c}$ and $T_{ex}$ dominate. We remind the reader here that the continuum temperature and excitation conditions are likely \textit{not} constant through the entire outer Galaxy, and thus a more precise treatment in the future will be warranted when more data are available.

\begin{figure*}
    \includegraphics[width=\textwidth]{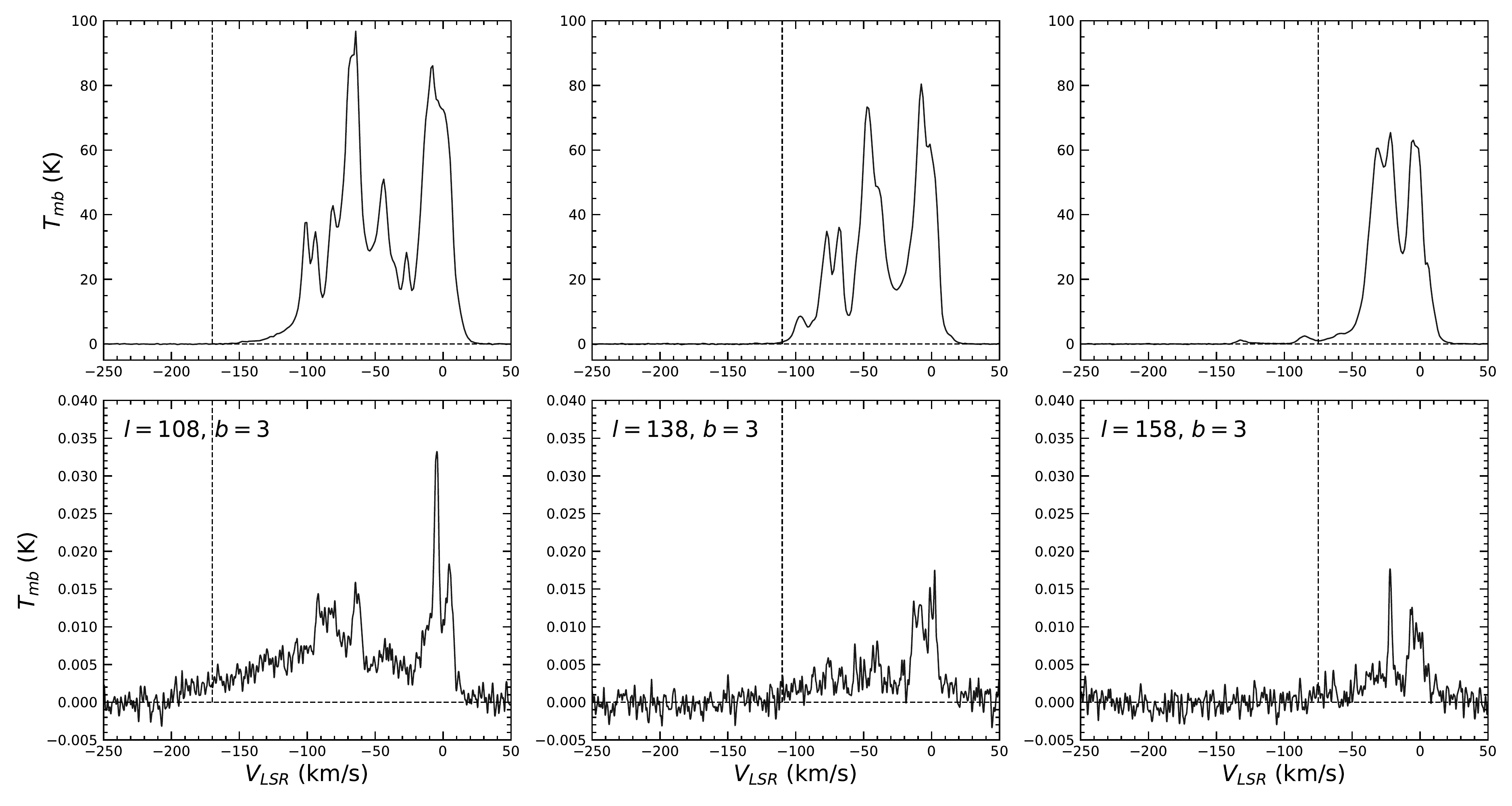}
    \caption{Left: The atomic (top) and molecular (bottom) content towards the sightline $l=108^{\circ}$, $b=3^{\circ}$ as traced by HI and OH with a 7.6' beam. Middle: The same at $l=138^{\circ}$, $b=3^{\circ}$. Right: The same at $l=158^{\circ}$, $b=3^{\circ}$. Notice that because $l$ is approaching $180^{\circ}$, the velocity of the emission begins to crowd towards 0 \kmps due to the geometry of the Galactic velocity field.}
    \label{fig:longitude}
\end{figure*}

\subsection{Column Density of HI} \label{HI}

An accurate measurement of the column density of HI is needed to calculate the presumed amount of \Htwo\ in the sightline from the gas content as traced by dust reddening. We estimate the optical depth of HI, $\tau$, as a constant value across the sightline by examining two absorption-emission pairs from \citet{Strasser2007Absorption} that we read from their online data: \textit{l} = 108.446\degr, \textit{b} = 4.0498\degr; and \textit{l} = 108.753\degr, \textit{b} = 2.5772\degr. Although $\tau$ changes along the line of sight, we take the median value from these sight lines as a constant for the calculations of the column density of HI at $\textit{l} = 108\degr$, $\textit{b} = 3\degr$ because they are close in proximity and representative of the diffuse ISM in this direction. The median across the line of sight in these two directions results in a $\tau$ = 0.103 $\pm$ 0.03, which we use as the $\tau$ in the following HI column density calculations.

We use Equation 5 in \citet{Dickey1982NeutralInstrument} to calculate the column density of HI using the LAB survey HI spectrum at l = 108\degr, b=3\degr: 

\begin{equation}
    \left<N(\mathrm{HI})\right> = C_{0} \int T_{mb}(v) \frac{\tau}{1 - e^{-\tau}} dv
\end{equation}

where $C_{0}$ = 1.82 $\times$ 10$^{18}$ $\frac{\mathrm{cm^{-2}}}{\mathrm{K}  \kmps}$. This calculation assumes an isothermal medium along the line of sight, with constant HI spin temperatures; which is likely somewhat incorrect because there are different environmental regions along the entire line of sight towards the outer Galaxy. However, because our estimate of the total gas content, $N_{H}$ is from dust IR emission, and hence does not contain radial velocity information, we cannot reliably decompose the HI and OH spectrum into their different components if the \Htwo\ content is of pertinent interest. The resulting HI column density is $N($H$)$ = (1 $\pm$ 0.3) $\times$ 10$^{22}$ cm$^{-2}$. The resulting N(OH)/N(H) ratio is $\sim$ a few $\times$ 10$^{-7}$.

The HI4PI server\footnote{https://www.astro.uni-bonn.de/hisurvey/AllSky\_gauss/} returns an HI column of $N($H$)$ = 9.58 $\times$ 10$^{21}$ cm$^{-2}$. Our column density differs (albeit, within error) because we restrain our column density calculation only to the Milky Way disk $(-150$ \kmps <  V$_{LSR}$ < $100$ \kmps$)$ and correct for $\tau$, whereas the HI4PI N(H) maps calculate their optically-thin column densities along their entire radial velocity range, $-600$ \kmps < $V_{LSR}$ < $600$ \kmps. This may include not only the Milky Way disk material, but also HI gas in the MW halo, intermediate and high velocity gas complexes, and extra-galactic HI objects.

\subsection{The OH Abundance Ratio}\label{XOH}

While several estimates of the N(OH)/N(H$_{2}$) abundance ratio exist for several Galactic environments \citep{Liszt1996GalacticSources,Nguyen2018Dust-GasISM,Rugel2018OHTHOR,Jacob2019FingerprintingDeconvolution}, it is unclear which value is appropriate to use for this sightline, as most of these estimates were calculated towards discrete clouds with absorption sources, and typically near the solar neighborhood. We estimate our own abundance ratio in this sightline by using dust reddening as a tracer for the total gas content, to see if it is consistent with the latest abundance ratio calculations from the THOR survey \citep{Rugel2018OHTHOR} and using CH as a calibration tracer \citep{Jacob2019FingerprintingDeconvolution}. We closely followed the prescription of this method as laid out in \citet{Nguyen2018Dust-GasISM}. Reddening is caused by dust grains through the absorption and scattering of light, defined as:

\begin{equation}
    E(B-V) = \frac{A_{V}}{R_{V}} = 1.086 \frac{\kappa}{R_{V}}r\mu m_{H} N_{H}
\end{equation}

\begin{figure*}[t]
        \centering
        \includegraphics[width=0.8\textwidth]{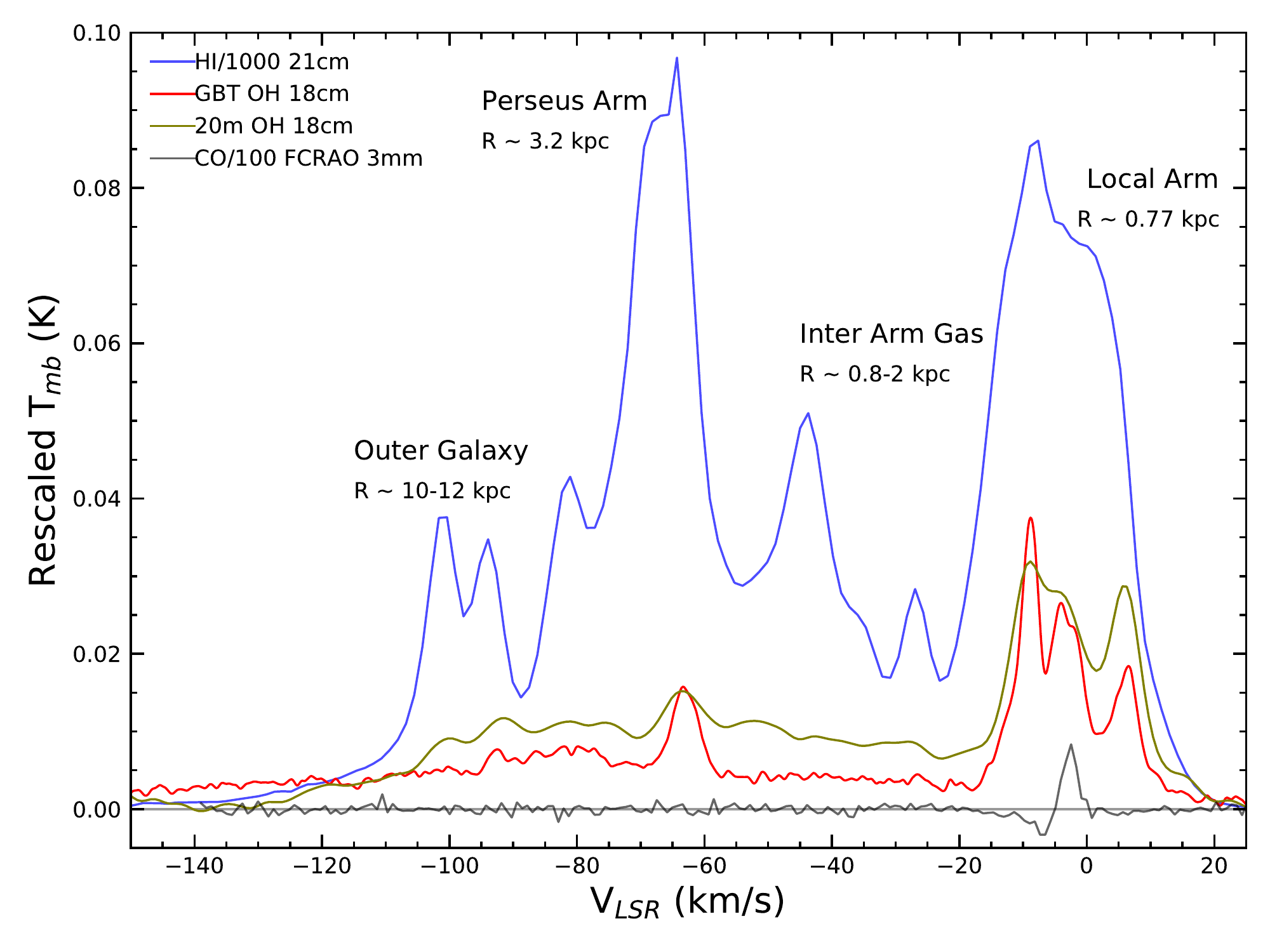}
    \caption{The observed 1667 and 1665 MHz OH broad emission toward the outer Galaxy ($l=108$, $b=3$), as observed by the 100m GBT and 20m telescope at the GBO. The dashed lines are the 3$\sigma$ (corresponding to 2mK for the 20m, 3mk for the 100m) statistical detection limit. The HI spectrum is from the LAB survey \citep{Kalberla2005TheHI}. The CO spectrum is from the FCRAO spectrum \citep{Heyer1998TheGalaxy}, kindly provided by M. Heyer.}
    \label{fig:PedestalAllSpectrum}
\end{figure*}

where $A_{V}$ is the dust extinction, $R_{V}$ is an empirical coefficient, $\kappa$ is the emissivity per cross section, $r$ is the dust-to-gas mass ratio, $\mu$ is the molecular weight (=2), $m_{H}$ is the weight of a hydrogen atom and $N_{H}$ is the total gas content, defined as: $N_{H} = N(H) + 2N(H_{2})$. $R_{V}$ is typically assumed to be 3.1 \citep{Cardelli1989TheExtinction}. While it has been shown that $R_{V}$ varies on Galactic scales, it does not vary significantly \citep{Schlafly2016THEWAY}. The SFD dustmap assumes a value of 3.1, which we use in this paper. 

In this analysis we assume that the total gas content is a linear combination of atomic and molecular hydrogen, $N_{H} = N(\mathrm{H}) + 2N(H_{2})$. We subtract the column density of atomic hydrogen and assume that the residual is molecular hydrogen. The uncertainties in this method include the uncertainty in the dust reddening, the uncertainty in the $N_{H}$/E(B-V) ratio, and the uncertainty in the optical depth, $\tau$, of the HI.

Using the opacity corrected HI column density from Sect. \ref{HI}, we calculate a column density of protons, N$_{H}$, of 1 $\times$ 10$^{22}$, and \Htwo\ of N(H$_{2}$) = 3 $\times$ 10$^{20}$ cm$^{-2}$. Using this to calculate an abundance ratio, N(OH)/N(H$_{2}$) = 1.5-4 $\times$ 10$^{-6}$, which is consistent the range of abundance ratios from the literature \citep{Nguyen2018Dust-GasISM, Rugel2018OHTHOR, Jacob2019FingerprintingDeconvolution}. Our calculations and the predictions from theory are also consistent with expectations for diffuse regions, $A_{v}$ < 0.1 mag \citep{Neufeld2016TheClouds}. 

\section{Discussion}\label{discussion}

\subsection{The large-scale Galactic structure of the OH Emission}\label{structure}

In 2019 we undertook a GBT survey in longitude and latitude to probe the structure of the broad OH emission under Project ID AGBT19A\_484. It was quickly discovered that the emission experiences the same expected velocity structure of the large-scale HI in the outer Galaxy. That is, the velocity distribution of the emission approaches zero with increasing $l$, as shown in Fig. \ref{fig:longitude}. This unambiguously implies that the broad OH profile is indeed a large-scale structure in the Galaxy that obeys the geometry of the velocity field \citep{Burton1985Leiden-GreenMaps.}. The HI and OH emission morphology appear to behave the same in velocity space, implying a cospatial nature of this phase of molecular gas with atomic gas in the outer galaxy. The HI associated with the Galactic disk is known to extend outward from the center farther than optically observed portions of the Galaxy such as stars, as well as known molecular components detected with CO. As such, this result extends the distance outward in the Galaxy in which molecular gas is found. At the same longitude and latitude, we demonstrate that this gas is by definition \textit{dark}, because it lacks a similar detection in a sensitive CO spectrum \citep{Heyer1998TheGalaxy}, see Fig. \ref{fig:PedestalAllSpectrum}.

While the longitude-velocity dependence is suggestive of a large-scale structure, we also checked for structure between $-5^{\circ} < b < 5^{\circ}$ towards $l = 105^{\circ}$ by stacking in latitude the data from AGBT15B\_004 in longitude. We were able to detect the broad OH emission to some level in every spectrum. It was apparent, however, that the OH emission is very small ($T_{mb}$ = 1 mK) at high latitudes, $b = -5^{\circ}$ and $b = 5^{\circ}$. This suggests that the feature has some scale height of $\sim$ 200 pc. In a future paper we will investigate this further. Additionally, it appears that the gas has a bi-modal distribution around $b = 1^{\circ}$. This may be because of the Galactic warp in this direction \citep{Yuan1973TheArms}, which warps the stellar and gas disk above the Galactic plane.

\subsection{The Implied Volume Density of the Diffuse Molecular Disk}\label{volumedensity}

Owing to Galactic rotation, the distance $s$ along the line of sight and the radial velocity $v$ at that point are approximately linearly related in the outer Galaxy. This suggests that we can estimate the $volume\ density\ n_{OH}$ at any point $s$ along the line of sight by differentiating equation \ref{eqn:coldens} with respect to distance $s$ along the line of sight to obtain:

\begin{eqnarray}
    n_s(OH) & \approx & C_{67} \left[\frac{T_{ex}^{67}}{T_{ex}^{67} - T_c(l\geq L)}\right] d/ds  \int \Delta T_b^{67}(v) dv \\
     & \approx & C_{67} \left[\frac{T_{ex}^{67}}{T_{ex}^{67} - T_c(l\geq L)}\right] \Delta T_b^{67}(v) dv/ds|_{s=S}
\end{eqnarray}

\noindent where $dv/ds$ is in units of \kmps/cm for the value of $C_{67}$ adopted here. To determine an approximate value for $dv/ds$ appropriate for any given Galactic longitude, we use the values of $S$ for a given value of $v$ returned by the VLBI ``Revised Kinematic Distance Calculator`` developed by the ``BeSSeL``  collaboration\footnote{http://bessel.vlbi-astrometry.org/home}; for additional background information see the review paper by \citet{Reid2014MicroarcsecondAstrometry}. This approach allows us to estimate values for $s(v)$ and $ds/dv$ which are less prone to local perturbations in the velocity field of the outer Galaxy. For instance, at $l=108.0\degr$, $b=+3.0\degr$, and e.g. at $v = -80$ \kmps (the approximate radial velocity of the outer arm of the Galaxy at this value of $l$) we obtain the linear approximation $s = a - bv$ with $a \approx 0$ and $b \approx 0.0875$ kpc/(\kmps) at a distance of $\approx 7.0$ kpc from the Sun. As an illustration, the volume density at this location is 1.8 $\times$ 10$^{-9}$ OH molecules \pcmcub per mK of 1667 MHz main-beam brightness temperature, assuming we are far enough in the outskirts of the Galaxy to set $T_c(s\geq S)=3$K (i.e. slightly above CMB alone), and choosing $T_{ex}^{67} = 4.1$K (a reasonable value between the ones found in \citet{Engelke2018OHW5} and \citet{Li2018WhereDMG}). Using the abundance ratio N(OH)/N(\Htwo) = 2 $\times$ 10$^{-6}$, the corresponding \Htwo\ volume density is $\sim$ 7 $\times$ 10$^{-3}$ \Htwo\ molecules \pcmcub at this location. By changing the excitation and continuum temperatures within reasonable assumptions ($\pm$ 2 K, for either), the volume density could be higher or lower by a factor of 10. This is consistent with the volume density of the  diffuse molecular gas phase suggested by \citet{Papadopoulos2002MolecularDistances}, and consistent with the average molecular volume density of the ISM found by \citet{Bohlin1978AII}.

\begin{figure*}[t]
        \centering
        \includegraphics[width=0.8\textwidth]{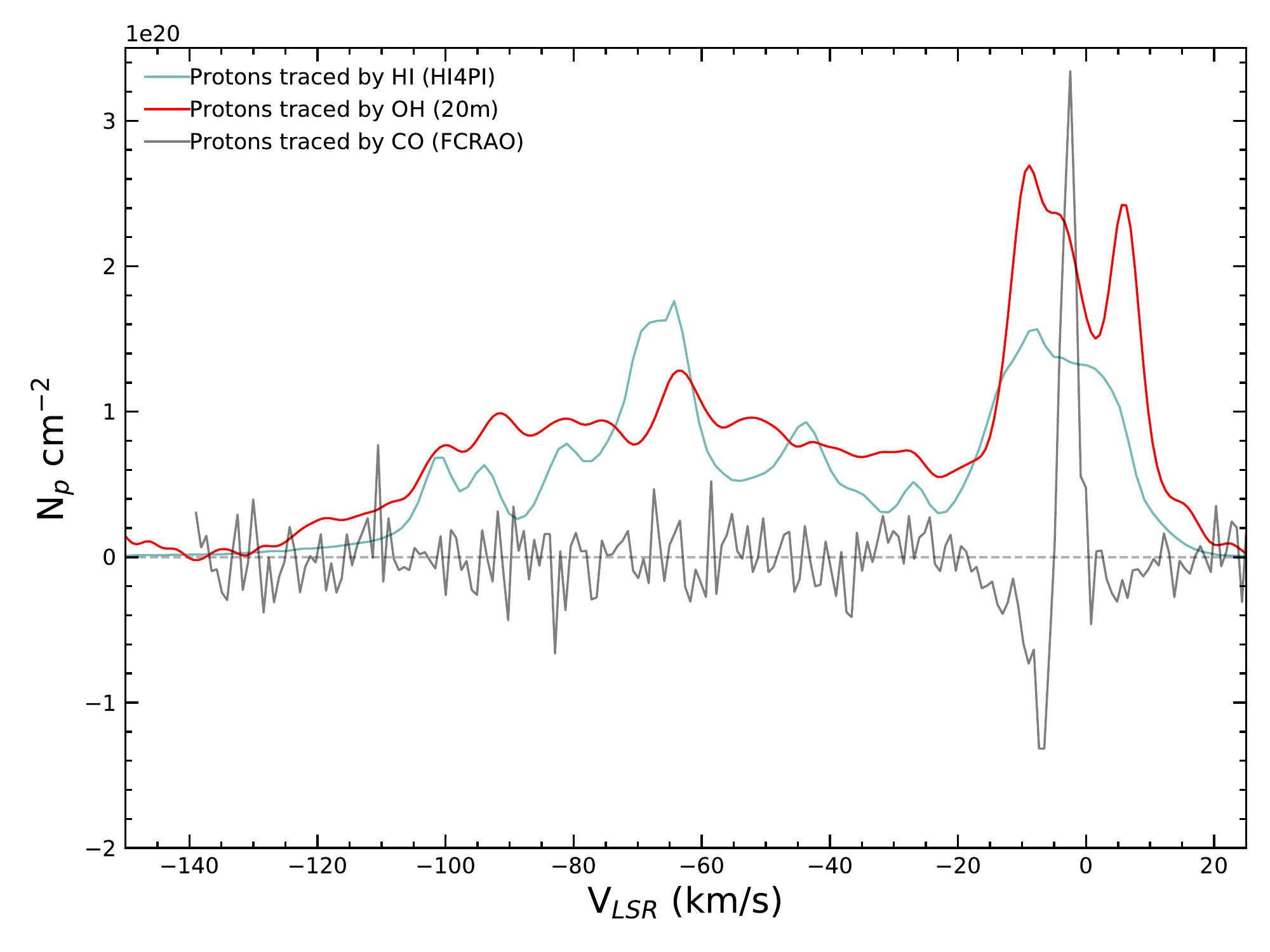}
    \caption{The observed HI, OH and CO spectrum from Fig. \ref{fig:PedestalAllSpectrum}, transformed into estimated number of protons traced by each tracer. Estimated number of protons for the atomic gas component is calculated directly from the HI signal; estimated number of protons for the molecular gas component is calculated from the OH signal using our assumed value of N(OH)/N($H_{2}$) of 2 $\times$ 10$^{-6}$, and from the CO signal using the X(CO) value 2 $\times$ 10$^{20}$ cm$^{-2}$ (K \kmps)$^{-1}$ from \citep{Bolatto2013TheFactor}. We see that diffuse molecular gas (as traced by OH) may be a component in mass equal to the atomic gas traced by HI, while much of the molecular gas appears to be CO-dark.}
    \label{fig:PedestalProton}
\end{figure*}

\subsection{Implications for Diffuse Molecular Gas in the Outer Galaxy}\label{H2}

An estimate of the \Htwo\ mass implied by the broad OH emission from the diffuse OH emission can be calculated using several reasonable assumptions. It is, however, important that we treat this number with caution because the uncertainties involved in this method amount to roughly an order of magnitude. Since we calibrate our OH/\Htwo\ abundance ratio for this sightline in an integrated fashion with dust reddening, we have no knowledge of the evolution of the OH/\Htwo\ abundance ratio in radial velocity (and hence, distance towards the Outer Galaxy). With the assumption of excitation and continuum temperature above, we move forward by transforming our column density of OH to \Htwo\ using the abundance ratio calculated in Sect. \ref{XOH}. The uncertainty in the abundance ratio is the dominant error in the following calculation. While we calculate an N(OH)/N(H$_{2}$), it is subject to all of the assumptions implicit in the dust reddening method, and we should also not be using one number for the abundance ratio for the sightline, as there are multiple environments in the spectrum. Nevertheless, if we use the derived abundance ratio, and the X(CO) number we can transform the T$_{mb}$ measurements of the HI, OH and CO spectrum in proton units. This exercise, displayed in Fig. \ref{fig:PedestalProton}, shows that diffuse molecular gas (as traced by OH) may be a component in mass equal to the atomic gas traced by HI in the outer Galaxy.

We use a simple geometric model of the outer Galaxy: a cylinder with a hole in the middle corresponding to the solar circle (R = 8 kpc). We calculate the volume of this cylinder as $V = \pi h (r_{1}^{2} - r_{2}^{2})$, where we take the height of the dark molecular disk as 200 pc, which is motivated by observations but will be further investigated in a future paper. In broad strokes, the z-distribution of OH was investigated at the ``Perseus Arm'' gas velocity ($V_{LSR} = -60$ \kmps), by fitting a Gaussian we were able to measure the scale height of the gas features. We take this as the scale height because it appears that the disk of emission peaks in surface brightness at this feature. We assume a line of sight disk length of R = 25 kpc, which is approximately the extent of the HI disk. The resulting volume of this hollow cylinder is 1 $\times$ 10$^{67}$ cm$^{3}$. Using a T$_{mb}$ = 4 mK, the number density is 7 \Htwo\ molecules per liter, the mass in the outer Galaxy can be calculated as: $M = 2 m_{p} V n_{\mathrm{H_{2}}}$ $\sim$ 1 $\times$ 10$^{8}$ M$_{\odot}$ The main uncertainties here are: the excitation temperature of OH and the abundance ratio of OH to \Htwo\, the abundance ratio being the dominant contribution to the uncertainty. 

The current literature values of OH/\Htwo\ cluster around $\sim$ 1 $\times$ 10$^{-7}$ \citep{Liszt1996GalacticSources,Nguyen2018Dust-GasISM,Rugel2018OHTHOR,Jacob2019FingerprintingDeconvolution}; a possible value of the mass using this ratio (which results in 150 \Htwo\ molecules per liter) is $\sim$ 2 $\times$ 10$^{9}$ M$_{\odot}$, which is about equal to current estimates of CO-traced \Htwo\ mass in the Galaxy at $\sim 1 \times 10^9 M_{\odot}$ \citep{Heyer2015MolecularWay}, and within an order of magnitude of the HI mass, at $\sim 8 \times 10^9 M_{\odot}$ \citep{Kalberla2009TheWay}.

If we use our calculated value for the abundance ratio for the lower limit of mass, and the literature value of the abundance ratio for the upper limit of the mass, the resulting range of molecular disk mass values is $\sim$ 10$^{8-9}$ M$_{\odot}$. The existence of a mass of dark \Htwo\ in our Galaxy equal to the CO-bright-traced \Htwo\ has been inferred from the indirect observations of $\gamma$-ray \citep{Grenier2005UnveilingNeighborhood} and dust \citep{PlanckCollaboration2011PlanckGalaxy}. We provide with this measurement concrete evidence of the existence of these diffuse molecules and, with the radial velocity distribution, the geometry of the diffuse molecular gas as a thick disk beyond the Solar circle. Note that the newly discovered diffuse molecular disk contains CO-dark molecular gas in addition to the previously reported clouds of CO-dark molecular gas found during our OH observing program in the arms of the Galaxy \citep{Allen2015The+1deg,Busch2019TheArm}. Moreover, it is interesting to examine to what extent the diffuse molecular disk might affect the total baryonic mass of the Galaxy. Using an estimate of stellar mass of $ \sim 6 \times 10^{10} M_{\odot}$ \citep{Licquia2015IMPROVEDMETA-ANALYSIS, McMillan2017TheWay} and an estimate of the known mass of the interstellar medium of $\sim 1 \times 10^{10} M_{\odot}$ \citep{Kalberla2009TheWay}, we roughly estimate that this newly discovered diffuse molecular disk could increase the baryonic mass estimate of the Galaxy by $\sim$ 0.1 $\%$  to $ \sim$ 2 $\%$. We anticipate that further sensitive surveys of this broad OH emission throughout the Galaxy will refine the measurement of the mass of the diffuse molecular disk in the future. 

The scale height of the diffuse OH emission corresponds extremely well with the thick diffuse molecular gas disk found in M51 in extended CO emission \citep{Pety2013TheGalaxy}. This comparison raises the question as to why the CO sightlines in the Milky Way are CO-dark, when the diffuse molecular gas in certain other galaxies others are not. One possible explanation for the detectable CO emission in M51 is the geometric effect of M51 being face-on. This orientation could result in enough radiative trapping in the z-direction for the emission from the diffuse component to be detectable at a sensitivity of $\sim$ 1 K \kmps. We show in \citet{Busch2019TheArm} that CO is subthermally excited below its critical density of 1000 cm$^{-3}$, and below this threshold the surface brightness of CO is heavily dependent on gas volume density. While the CO emission in \citet{Pety2013TheGalaxy} appears to come from diffuse molecular gas of n $\sim$ 300-500 cm$^{-3}$, the OH emission may yet still be tracing an even more diffuse component in addition to the CO-traced diffuse molecular gas because its critical density is n $\sim$ 1 cm$^{-3}$. A future study intended to search for OH emission from diffuse disks in external galaxies known to contain CO-bright disks, as well as several that do not contain detected CO disks, could provide more information about the differences between conditions in the disks of these galaxies and the properties that lead to a diffuse disk being CO-bright or CO-dark. Such a study could also demonstrate whether OH traces a larger total mass of diffuse molecular gas in these disks even for galaxies that already contain a detectable disk in CO.

\citet{Papadopoulos2002MolecularDistances} lays out a theoretical basis for both \Htwo\ formation and shielding at large Galactic distances. They argue that both the ingredients (HI and dust grains) and environmental conditions exist for \Htwo\ formation well inside an HI disk. The existence of abundant dark \Htwo may also shed light on the puzzling star formation behaviours in HI dominated regions \citep{Krumholz2013TheGalaxies}. Star formation correlates with \Htwo\ down to the lowest column densities that can be detected by CO, but then also begins to correlate with HI in the outer parts of the Galaxy. If there were a thick disk of diffuse molecular gas in the outer Galaxy, this may help explain the star formation in the HI dominated portions of the outer Galaxy, as opposed to alternative explanations such as high-velocity runaway stars \citep{Andersson2021RunawayOutskirts}.

Modeling and observations in future papers will have to study the distinction between the OH and CO-traced diffuse molecular gas disks in order to fully understand the transition zone between CO-bright and CO-dark molecular gas.

\section{Conclusions}\label{conclusions}

We have discovered an extremely broad OH emission profile towards the outer Galaxy that is cospatial with the HI velocity distribution. We have convinced ourselves that it is a real astronomical signal by observing the feature in: both 1667 and 1665 MHz OH lines, two different telescopes with vastly different beams and associated instrumental effects, and using three different data pipelines with which to reduce the data.

This result indicates that the molecular disk of the Galaxy is significantly more extended than previously suspected. We believe that the geometry of the molecular gas traced by the emission is in the form of an extremely diffuse disk, with a thickness observed in the Galactic latitude range $-5^{\circ} < b < 5^{\circ}$. The signal is indicative of large-scale structure because it obeys the velocity field of the Galaxy. 

A conceptual model of the outer Galaxy using our derived value of the volume density of $\sim$ tens of H$_2$ molecules per liter (7-70 $\times$ 10$^{-3}$ cm$^{-3}$) results in a molecular mass value for the dark molecular disk of $\sim$ 10$^{8-9}$ M$_{\odot}$, with the order of magnitude uncertainty resulting from the adopted range of values of the abundance ratio, N(OH)/N(H$_{2}$).

We plan multiple follow up surveys to probe the latitude and longitude dependence of the OH emission, and to characterize the structure of the diffuse molecular gas using OH emission. We hope this discovery will motivate similar searches for extremely diffuse molecular gas using dark molecular gas tracers on a large, Galactic scale. We anticipate there is much to learn about Galactic structure by using an optically thin, widely observed molecular gas tracer in the Galaxy.

\pagebreak

\acknowledgments

We would like to dedicate this discovery to Ronald J. Allen, who passed away during the drafting of this manuscript. He was an amazing mentor, a brilliant astronomer, and a good friend to many in our profession. He led much of the early research in tracing the large scale structure of diffuse molecular gas through blind, sensitive emission surveys of Galactic OH, which this work was built upon. We will miss him dearly.

We are grateful to the scientific staff and telescope operators at the Green Bank Observatory for their advice and assistance with the operation of the GBT, in particular Karen O'Neil, Jay Lockman, Toney Minter, Ron Maddalena, and Amanda Kepley, and for the development and support of the GBTIDL data analysis software, especially Jim Braatz and Bob Garwood. 

We are especially grateful to Frank Ghigo of Green Bank Observatory, for a tremendous investment of his time and skills in the required maintenance of the 20m telescope and the development of a data pipeline for the observations required for this paper.

We are grateful to the anonymous referee, whose comments improved the contents of this paper.

Michael P. Busch is grateful to Josh Peek, Claire Murray, Susan Kassin, Christine Chen, Tobias Marriage, David Neufeld, Colin Norman, and Nadia Zakamska for their help and guidance through a difficult time in his graduate school career after RJA's passing.

MPB is supported by a National Science Foundation Graduate Research Fellowship under grant No. 1746891. MPB thanks the LSSTc Data Science Fellowship Program, which is funded by LSSTc, NSF Cybertraining Grant No. 1829740, the Brinson Foundation, and the Moore Foundation; their participation in the program has benefited this work. This research has been supported by the Director's Research Fund at STScI.

\software{\textit{Astropy} \citep{TheAstropyCollaboration2013Astropy:Astronomy}, \textit{dustmaps} \citep{M.Green2018Dustmaps:Dust}, \textit{GBTIDL} \citep{Garwood2006GBTIDL:Data}}

\bibliographystyle{apj}

\end{document}

%% file: mycommands.tex
\newcommand{\I}{\protect\small I \normalsize $\!\!$}

\newcommand{\Htwo}{H$_{2}$}

\newcommand{\twCO}{$^{12}$CO}

\newcommand{\pcmcub}{\mbox{${\rm cm^{-3}}$}}

\newcommand{\kmps}{\mbox{${\rm km\;s^{-1}}$}}

\newcommand{\pcmsq}{\mbox{${\rm cm^{-2}}$}}

\newcommand{\lsim}{\mbox{$\mathrel{\vcenter{\hbox{\ooalign{\raise3pt\hbox{$<$}\crcr \lower3pt\hbox{$\sim$}}}}}$}}
\newcommand{\gsim}{\mbox{$\mathrel{\vcenter{\hbox{\ooalign{\raise3pt\hbox{$>$}\crcr \lower3pt\hbox{$\sim$}}}}}$}}